\documentclass[aip, jcp, amsmath, amssymb, reprint]{revtex4-1}

\usepackage{graphicx}
\usepackage{dcolumn}
\usepackage{bm}
\usepackage{color}

\begin{document}


\title{Lattice model of linear telechelic polymer melts. II. Influence of chain stiffness on basic thermodynamic properties}

\author{Wen-Sheng Xu}
\email{wsxu@uchicago.edu}
\affiliation{James Franck Institute, The University of Chicago, Chicago, Illinois 60637, USA}

\author{Karl F. Freed}
\email{freed@uchicago.edu}
\affiliation{James Franck Institute, The University of Chicago, Chicago, Illinois 60637, USA}
\affiliation{Department of Chemistry, The University of Chicago, Chicago, Illinois 60637, USA}

\date{\today}

\begin{abstract}
The	lattice cluster theory (LCT) for semiflexible linear telechelic melts, developed in paper I, is applied to examine the influence of chain stiffness on the average degree of self-assembly and the basic thermodynamic properties of linear telechelic polymer melts. Our calculations imply that chain stiffness promotes self-assembly of linear telechelic polymer melts that assemble on cooling when either polymer volume fraction $\phi$ or temperature $T$ is high, but opposes self-assembly when both $\phi$ and $T$ are sufficiently low. This allows us to identify a boundary line in the $\phi$-$T$ plane that separates two regions of qualitatively different influence of chain stiffness on self-assembly. The enthalpy and entropy of self-assembly are usually treated as adjustable parameters in classical Flory-Huggins type theories for the equilibrium self-assembly of polymers, but they are demonstrated here to strongly depend on chain stiffness. Moreover, illustrative calculations for the dependence of the entropy density of linear telechelic polymer melts on chain stiffness demonstrate the importance of including semiflexibility within the LCT when exploring the nature of glass formation in models of linear telechelic polymer melts.
\end{abstract}



\maketitle

\section{Introduction}

Despite the increasing scientific interest and technological importance of telechelic polymers, a molecular theory that predicts the relation between the monomer molecular structure and the thermodynamic properties in such systems has been slow to develop. Prior theories~\cite{Mac_28_1066, Mac_28_7879, JCP_110_1781, Mac_33_1425, Mac_33_1443, JCP_119_6916, Lan_20_7860, JPSB_45_3285, JCP_131_144906, JPCB_114_12298} of self-assembly in telechelic polymers traditionally employ highly coarse grained models that represent the assembling molecular species as a structureless entity and hence, cannot address the influence of local molecular structure on the self-assembly of telechelic polymers. Dudowicz and Freed~\cite{JCP_136_064902} have recently reformulated the lattice cluster theory (LCT) for the thermodynamics of polymer systems to treat strongly interacting, self-assembling polymers composed of fully flexible linear telechelic chains. Because the LCT employs an intermediate level of coarse grained models that retain minimal aspects of monomer molecular structure and interactions in telechelic polymers, the theory enables establishing the relation between the molecular structure dependent interaction parameters of the model and the thermodynamic properties of these complex fluids.~\cite{JCP_136_064903, JCP_136_194902} Therefore, the LCT provides a potential framework for obtaining information that is useful for designing telechelic polymer materials. 

Paper I of this series~\cite{Paper1} further extends the LCT for the thermodynamics of linear telechelic polymers to include a description of chain semiflexibility. The significance of this extension is at least two-fold. Firstly, computer simulations~\cite{JCP_110_6039, Mac_47_4118, Mac_47_6946} indicate that chain semiflexibility strongly affects the self-assembly process and structural properties of telechelic polymers. The LCT for semiflexible telechelic polymers provides a theoretical tool for investigating the influence of chain stiffness on the thermodynamic properties of self-assembling telechelic polymers and hence, offers the possibility of better understanding experiments and simulations of telechelic polymers. Secondly, chain semiflexibility is crucial for exploring the behavior of glass formation in polymers within the generalized entropy theory (GET),~\cite{ACP_137_125} a theory of polymer glass formation that merges the LCT for the thermodynamics of semiflexible polymers~\cite{ACP_103_335} with the Adam-Gibbs (AG) relation between the structural relaxation time and the configurational entropy,~\cite{JCP_43_139, JCP_141_141102} because models of fully flexible polymer chains fail to exhibit the characteristic glassy behavior in the GET, as noted in Ref.~\citenum{JCP_140_244905}. 

The previous work~\cite{JCP_136_064903, JCP_136_194902} analyzes the influence of various molecular parameters on the thermodynamic properties only for fully flexible linear telechelic polymers. The influence of chain semiflexibility upon the thermodynamics of self-assembling telechelic polymers is investigated using the theoretical advances of the LCT in paper I,~\cite{Paper1} to explore the relation between molecular factors and thermodynamic properties in equilibrium self-assembling semiflexible telechelic polymers, and in particular, in semiflexible linear telechelic polymers.

Section II provides a brief summary of the LCT for semiflexible linear telechelic polymers. Section III begins by examining the influence of chain stiffness on the average degree of self-assembly of linear telechelic polymers that assemble on cooling. Our calculations demonstrate that chain stiffness can either promote or oppose the self-assembly of linear telechelic polymer melts, depending on the polymer volume fraction and the temperature. Specifically, while chain stiffness promotes self-assembly of linear telechelic polymer melts when either polymer volume fraction $\phi$ or temperature $T$ is high, self-assembly is suppressed by chain stiffness when both $\phi$ and $T$ are sufficiently low. This allows us to identify a boundary line in the $\phi$-$T$ plane, which separates two regions in which the influence of chain stiffness on the self-assembly is qualitatively different. We further show that the boundary line is independent of the sticky interaction energy but changes with bending energy, molecular weight, and van der Waals interaction energy. Section III then presents calculations of the enthalpy and entropy of self-assembly that are usually treated as adjustable parameters in classical Flory-Huggins (FH) type theories for the equilibrium self-assembly of polymers (e.g., see Refs.~\citenum{JCP_119_12645, JCP_130_164905, JCP_130_224906, JCP_136_244904}) but that are shown here to depend strongly on chain stiffness. Section III further presents calculations for the influence of chain stiffness on the entropy density of linear telechelic polymer melts, thereby illustrating the importance of including semiflexibility within the LCT for exploring the nature of glass formation in models of linear telechelic polymer melts.

\section{Lattice cluster theory for semiflexible linear telechelic polymer melts}

\begin{figure}[tb]
 \centering
 \includegraphics[angle=0,width=0.45\textwidth]{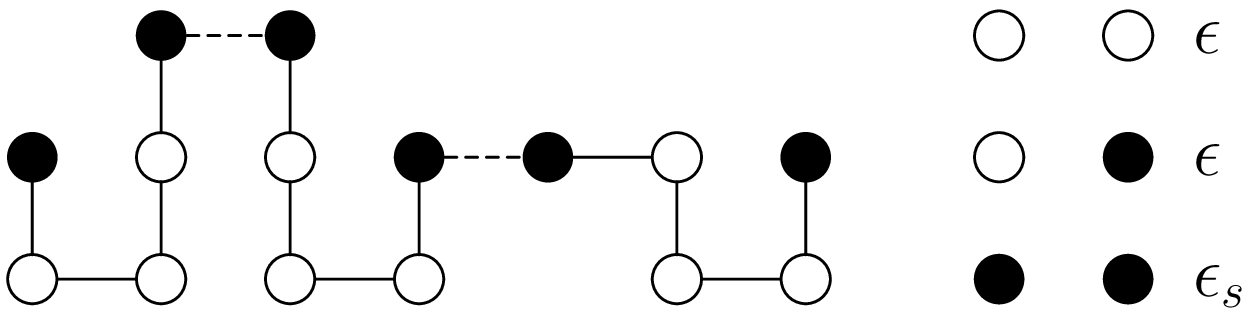}
 \caption{Schematic illustration of the lattice model for a self-assembled linear telechelic cluster composed of three short chains with $M=5$ united atom groups in each chain. Solid circles designate the chain's ends that can participate in sticky interactions, while open circles denote the united atom groups in the chain interior. Solid lines represent the chemical bonds between two united atom groups within the same chain, while dashed lines denote the sticky physical bonds between two stickers. The model prescribes two different nearest neighbor interaction energies $\epsilon$ and $\epsilon_s$  for ordinary and sticky-sticky interactions, respectively.}
\end{figure}

We briefly review the basic characteristics of the LCT for semiflexible linear telechelic polymer melts to provide necessary definitions, notation, and description of the model; more details can be found in paper I.~\cite{Paper1} 

The system consists of $m$ semiflexible linear chains that are located on a simple cubic lattice with lattice coordination number equal to $z=6$ and with $N_l$ total lattice sites. Since the system is treated as a compressible melt, the lattice also contains empty sites whose volume fraction is determined from the equation of state for a system at constant temperature and pressure. The volume fraction of polymer chains is defined by $\phi=mM/N_l$, where $M$ is the number of united atom groups per chain. In order to model telechelic chains, each chain's end segments (called ``stickers'' and represented as solid circles in Fig. 1) are distinguished from the other united atom groups lying in the chain interior (depicted by open circles in Fig. 1). Two stickers are allowed to form a sticky ``bond'' (pictured as dashed lines in Fig. 1) and interact with an attractive sticky interaction energy $\epsilon_s$ when they are located on nearest neighbor lattice sites.~\cite{JCP_136_064902} A negative $\epsilon_s$ implies that polymeric clusters form upon cooling. This sticky energy $\epsilon_s$ may greatly exceed the microscopic van der Waals interaction energy $\epsilon$, a parameter that describes the attractive interaction strength between two non-stickers as well as between a sticker and a non-sticker (see Fig. 1) and that is conventionally treated as positive in the original LCT.~\cite{Mac_24_5076, JCP_87_5534} The current LCT assumes that the stickers are mono-functional (i.e., the stickers at each of the ends of the telechelics can only participate in one sticky interaction) and that both cyclic and linear associative clusters may form, in accord with the previous work~\cite{ JCP_136_064902, JCP_136_244904} and the analysis of Jacobson and Stockmayer.~\cite{ JCP_18_1600} Following the previous treatment in the LCT,~\cite{ACP_103_335} chain semiflexibility is treated by introducing a bending energy penalty $E_b$ whenever a pair of consecutive bonds from a single telechelic chain lies along orthogonal directions. Because the sticky bonds are treated as being fully flexible, the theory leads to a model where the associated chains are monomeric rods with fully freely rotating junctions at the sticky bonds in the limit $E_b\rightarrow \infty$. This feature prevents the occurrence of crystallization in rigid linear telechelic polymers, and hence, may be important for exploring glass formation in such systems.

As derived in paper I,~\cite{Paper1} the Helmholtz free energy $f$ per lattice site of a semiflexible telechelic polymer melt is the sum of the free energy $f_o$ of the hypothetical reference system in the absence of sticky interactions and the free energy contribution $f_s$ arising from the sticky interactions,
\begin{equation}
\beta f=\beta f_o+\beta f_s,
\end{equation}
where $\beta=1/(k_BT)$ with $k_B$ being Boltzmann's constant and $T$ designating the absolute temperature. The free energy $f_o$ of the hypothetical reference system is identical to the system under consideration, except that the sticky interaction is absent. Thus, the free energy $f_o$ of the reference system is independent of $\epsilon_s$ and its derivation is described elsewhere in detail.~\cite{ACP_103_335, JCP_141_044909} 

The sticky contribution $f_s$ is derived by first defining the density $y$ of sticky bonds (i.e., the ratio of the number of sticky bonds in the system to the total number of lattice sites) and emerges as
\begin{equation}
\beta f_s=\beta f_s^{mf}-\sum_{i=1}^4Y_iy^i,
\end{equation}
and
\begin{eqnarray}
\beta f_s^{mf}=&&-\phi x\ln(\phi x)+(\phi x-2y)\ln(\phi x-2y)\nonumber\\
&&
+y\left[1+\ln\left(\frac{2y}{z}\right)+\beta\epsilon_s\right],
\end{eqnarray}
where $x=2/M$ is the fraction of stickers in a single chain, and $Y_i$ $(i=1,...,4)$ are corrections to the zeroth-order mean-field free energy $\beta f_s^{mf}$ arising from short range correlations up to the scale of four consecutive bonds. Explicit expressions for $Y_i$ $(i=1,...,4)$ are provided in paper I.~\cite{Paper1} The variable $y$ in Eqs. (2) and (3)  is determined by the maximum term method,~\cite{JCP_136_064902} i.e., by applying the condition,
\begin{eqnarray}
\left. \frac{\partial (\beta f_s)}{\partial y}\right|_{N_l, T, \phi}=0.
\end{eqnarray}
The solution of Eq. (4), i.e., the concentration $y^{\ast}$ of the sticky bonds, is then substituted into Eqs. (1-3), leading to the final expression for the free energy $f$ of a semiflexible linear telechelic melt, 
\begin{eqnarray}
\beta f=&&\beta f_o-\phi x\ln(\phi x)+(\phi x-2y^{\ast})\ln(\phi x-2y^{\ast})\nonumber\\
&&
+y^{\ast}\left[1+\ln\left(\frac{2y^{\ast}}{z}\right)+\beta\epsilon_s\right]-\sum_{i=1}^4Y_i(y^{\ast})^i,
\end{eqnarray}
which depends on all molecular and thermodynamic parameters including $T$, $\phi$, $M$, $\epsilon$, $E_b$, and $\epsilon_s$. Notice that the volume fraction of the active stickers (i.e., those participating in sticky interactions) is simply $2y^{\ast}$ and that the upper limit for $y^{\ast}$ is $y_{max}^{\ast}=\phi/M$, because the chain's ends are each assumed to be mono-functional in the present model.

While the present paper discusses the model and results with reference to compressible melts, the mathematical equivalence between the excess thermodynamic properties of a compressible melt and an incompressible solution enables describing both types of systems. The model for an incompressible solution consists of one-bead solvent molecules replacing the empty lattice sites.~\cite{JCP_136_064902} The free energy of a compressible polymer melt is isomorphic to that of an incompressible polymer solution, with the $\epsilon$ parameter being replaced by the exchange energy $\epsilon_{\text{ex}}=\epsilon_{pp}+\epsilon_{ss}-2\epsilon_{ps}$, where $\epsilon_{pp}$, $\epsilon_{ss}$ and $\epsilon_{ps}$ represent the strengths of the nearest neighbor interaction between two polymer segments, two solvent molecules and a polymer segment and a solvent molecule, respectively. 

\section{Results and discussion}

This section examines the influence of chain stiffness on the average degree of self-assembly and the enthalpy and entropy of self-assembly of linear telechelic polymer melts. This section also demonstrates the importance of including chain semiflexibility within the LCT for exploring glass formation in the model of linear telechelic polymer melts.

\subsection{Influence of chain stiffness on the average degree of self-assembly}

\begin{figure}[tb]
 \centering
 \includegraphics[angle=0,width=0.45\textwidth]{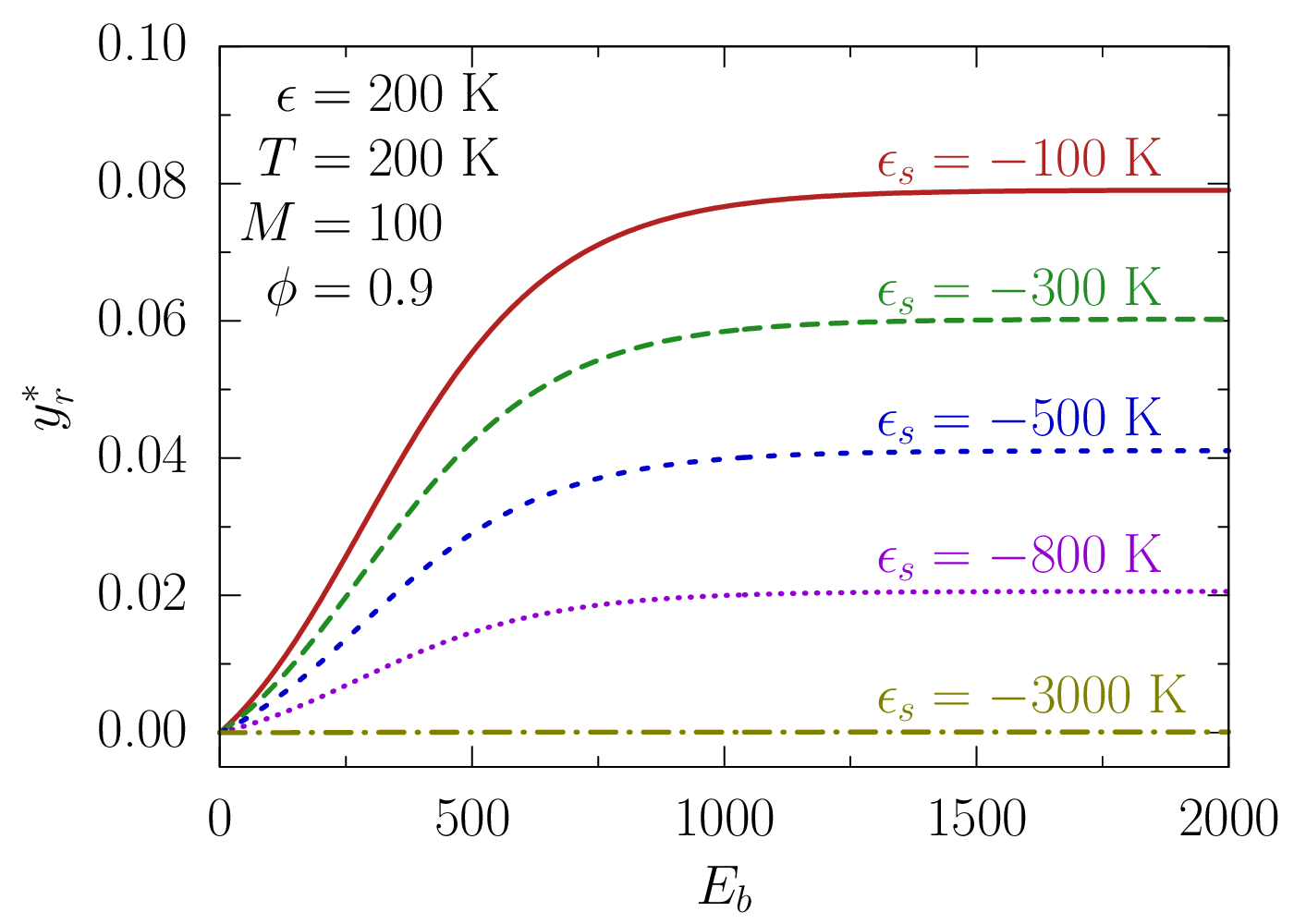}
 \caption{$y_r^{\ast}$ as a function of bending energy $E_b$ for various sticky interaction energies $\epsilon_s$. The computations are performed for a melt of linear telechelic chains, where the polymer volume fraction is $\phi=0.9$, the molecular weight is $M=100$, and the van der Waals interaction energy is $\epsilon=100$ K. The temperature is fixed at $T=200$ K.}
\end{figure}

\begin{figure}[tb]
 \centering
 \includegraphics[angle=0,width=0.45\textwidth]{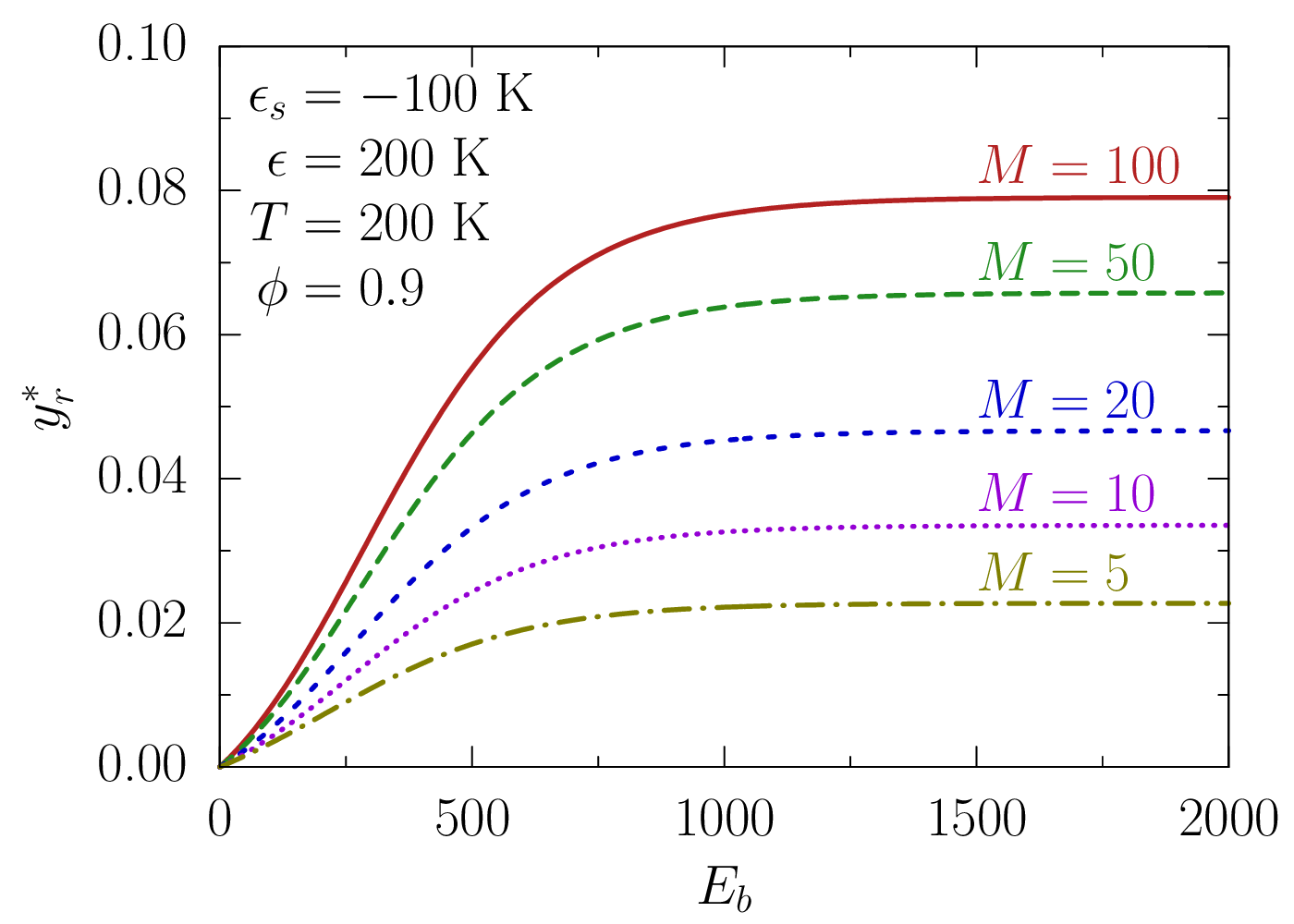}
 \caption{$y_r^{\ast}$ as a function of bending energy $E_b$ for various molecular weights $M$. The computations are performed for a melt of linear telechelic chains, where the polymer volume fraction is $\phi=0.9$, the van der Waals interaction energy is $\epsilon=200$ K, and the sticky interaction energy is $\epsilon_s=-100$ K. The temperature is fixed at $T=200$ K.}
\end{figure}

\begin{figure}[tb]
 \centering
 \includegraphics[angle=0,width=0.45\textwidth]{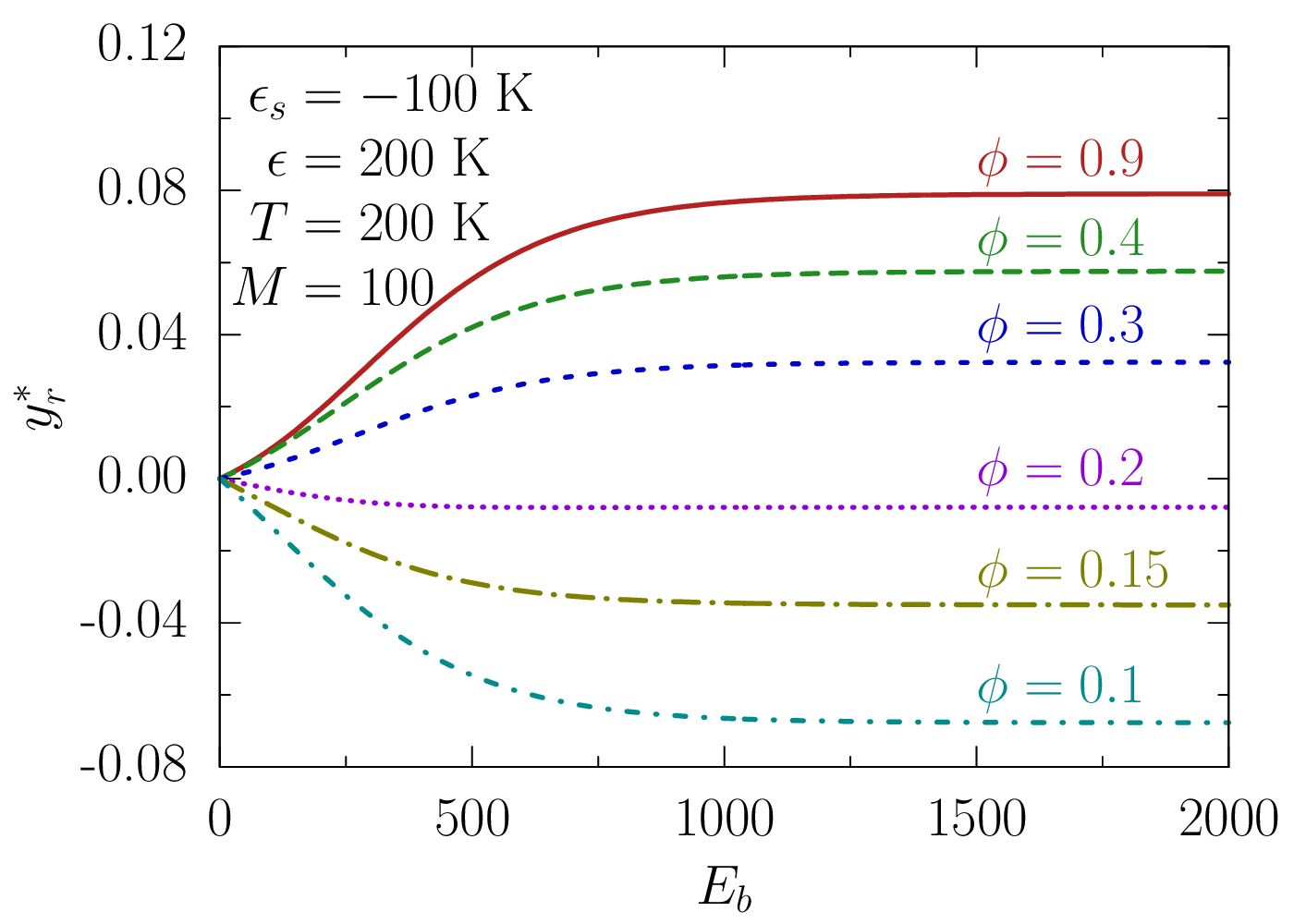}
 \caption{$y_r^{\ast}$ as a function of bending energy $E_b$ for various polymer volume fractions $\phi$ for a melt comprising linear telechelic chains, where the molecular weight is $M=100$, the van der Waals interaction energy is $\epsilon=200$ K,  the sticky interaction energy is $\epsilon_s=-100$ K, and the temperature is fixed at $T=200$ K.}
\end{figure}

\begin{figure}[tb]
 \centering
 \includegraphics[angle=0,width=0.45\textwidth]{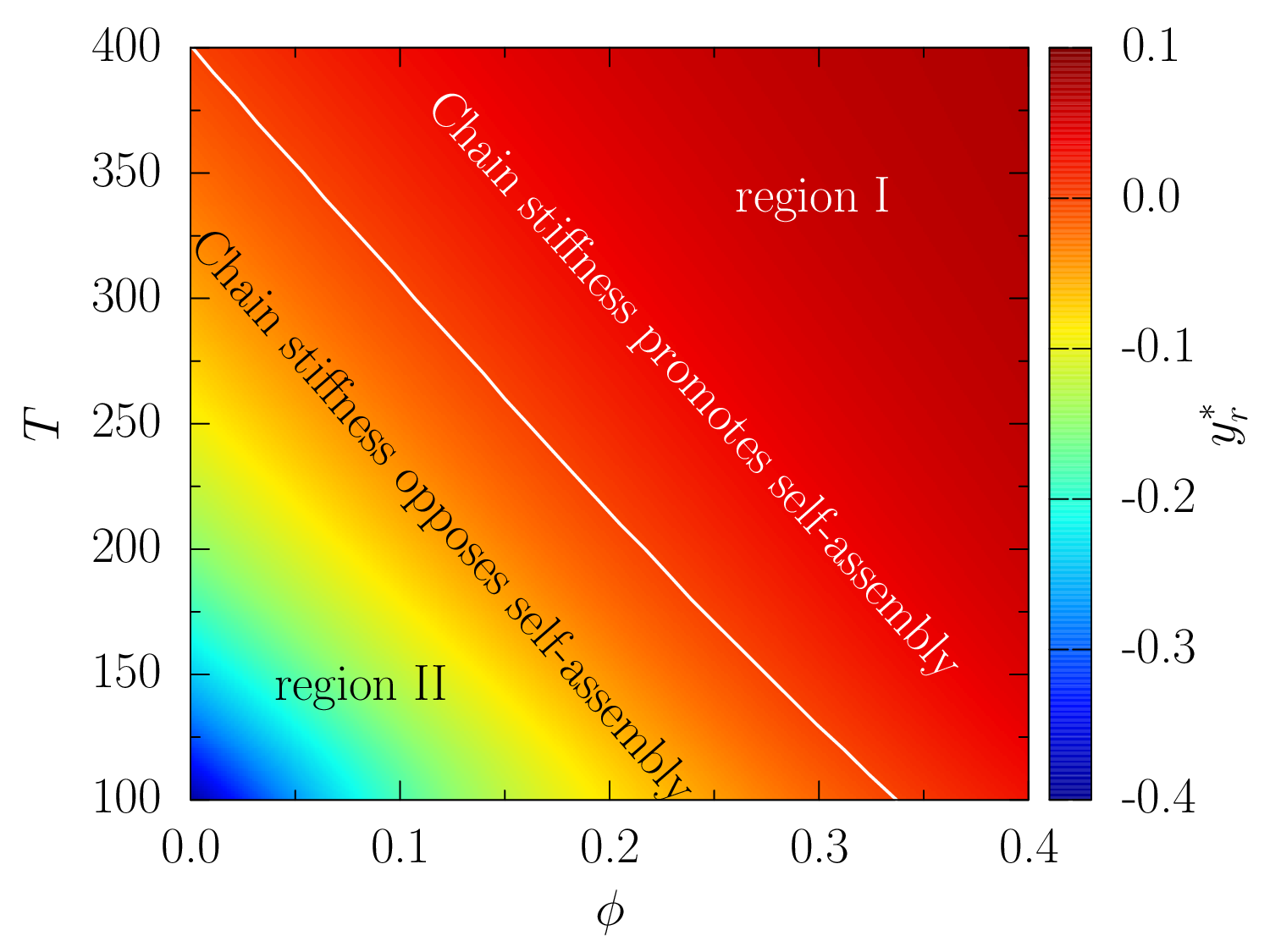}
 \caption{Contour plot of $y_r^{\ast}$ in the $\phi$-$T$ plane. The computations are performed for a melt of linear telechelic chains, where the molecular weight is $M=100$, the van der Waals interaction energy is $\epsilon=200$ K, the sticky interaction energy is $\epsilon_s=-100$ K, and the bending energy is $E_b=2000$ K. The solid line denotes the boundary marking the states with $y^{\ast}=0$.}
\end{figure}

\begin{figure*}[tb]
 \centering
 \includegraphics[angle=0,width=0.85\textwidth]{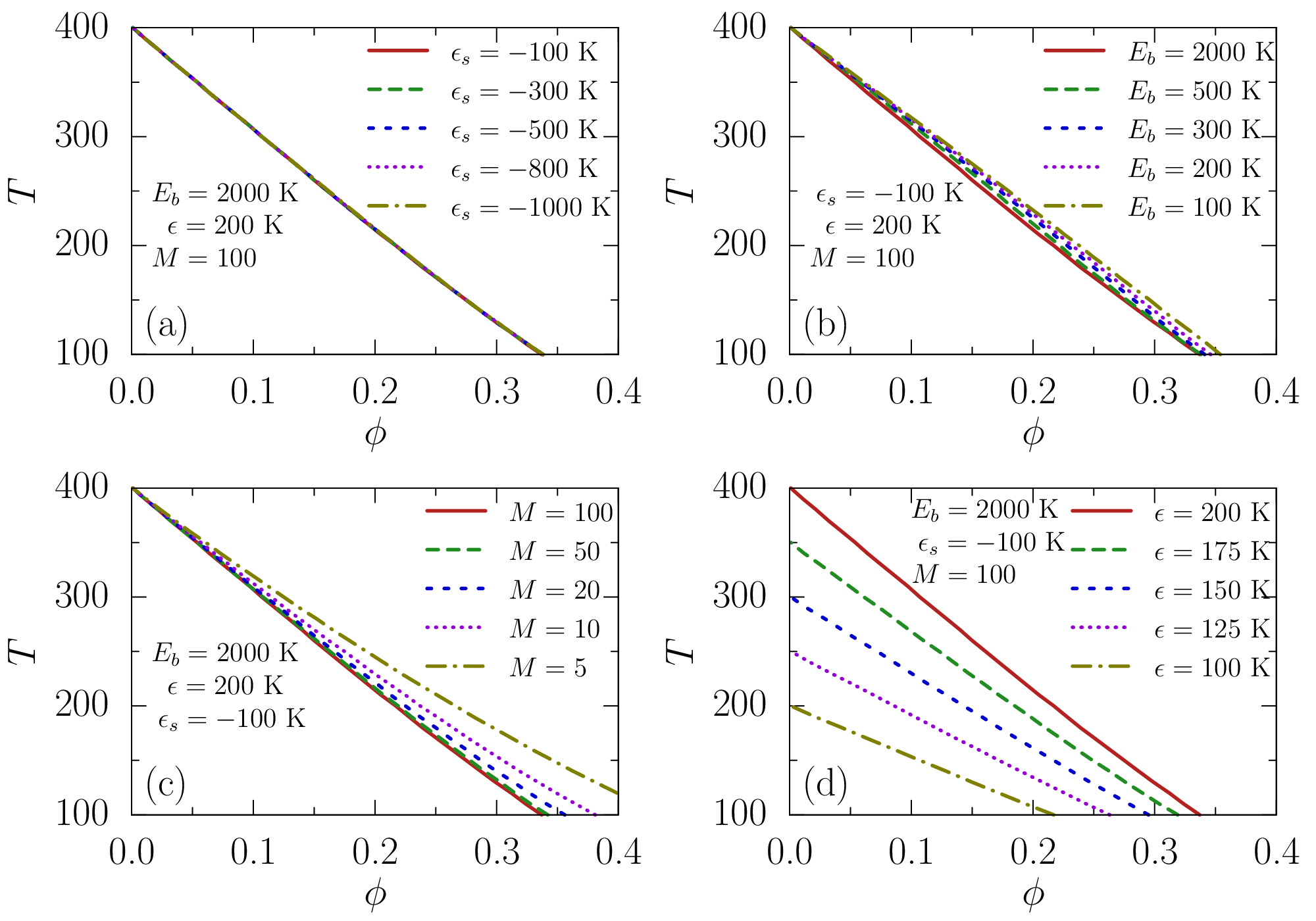}
 \caption{Dependence of the boundary line in the $\phi$-$T$ plane on (a) the sticky interaction energy $\epsilon_s$, (b) the bending energy $E_b$, (c) the molecular weight $M$, and (d) the van der Waals interaction energy $\epsilon$.}
\end{figure*}

An understanding of the self-assembly of telechelic systems is crucial for many practical applications.~\cite{POC_7_289, Book_Goodman, Polymer_45_3527, Nature_453_171} As explained in Sec. II and paper I,~\cite{Paper1} the concentration $y^{\ast}$ of the sticky bonds measures the volume fraction of the active stickers in the LCT for linear telechelic polymer melts. Hence, $y^{\ast}$ directly reflects the average degree of self-assembly and other important quantities such as the order parameter $\Phi$ and the transition temperature $T_p$ of self-assembly.~\cite{JCP_136_194902} (The average degree $<N>$ of self-assembly is more precisely defined by $<N>\approx1/(1-\Phi)$ with $\Phi=y^*/y^*_{max}$; see Ref.~\citenum{JCP_136_194902} for more details.) We thus focus on the general trends exhibited by the variation of $y^{\ast}$ with bending energy. Because the dependence of $y^{\ast}$ on various molecular and thermodynamic parameters, such as $T$, $\phi$, $M$, $\epsilon$, and $\epsilon_s$, has been extensively examined in Refs.~\citenum{JCP_136_064903, JCP_136_194902} for the model of fully flexible linear telechelic polymers, the computations presented in this subsection are devoted to understanding the role of chain stiffness in the self-assembly of linear telechelic polymer melts. As in previous work,~\cite{JCP_136_064903, JCP_136_194902} the lattice coordination number is taken as $z=6$ in all our computations.

Because the quantitative effect of $E_b$ on $y^{\ast}$ is generally found to be small numerically, we introduce the ratio $y_r^{\ast}$ that measures the relative change of $y^{\ast}$ with increasing $E_b$,
\begin{equation}
y_r^{\ast}=\frac{y^{\ast}-y_0^{\ast}}{y_0^{\ast}},
\end{equation}
where $y_0^{\ast}$ is the value of $y^{\ast}$ for $E_b=0$ (i.e., the fully flexible case). Apparently, a positive or negative $y_r^{\ast}$ implies that chain stiffness promotes or opposes self-assembly, respectively. 

Figure 2 displays $y_r^{\ast}$ as a function of $E_b$ for various sticky interaction energies $\epsilon_s$, when the other parameters are held constant. $y_r^{\ast}$ first increases with $E_b$ and then saturates at sufficiently large $E_b$ for small $|\epsilon_s|$. Moreover, the quantitative dependence on $E_b$ weakens as $|\epsilon_s|$ increases and even disappears when $|\epsilon_s|$ is sufficiently large (e.g., $|\epsilon_s|=3000$ K), and the saturated value for $y_r^{\ast}$ decreases with increasing $|\epsilon_s|$. This is understandable because $y^{\ast}$ is shown to increase with $|\epsilon_s|$ and reach its upper limit (i.e., $\phi/M$) for large $|\epsilon_s|$.~\cite{JCP_136_064903} Clearly, $y^{\ast}$ fails to elevate with increasing $E_b$ after $y^{\ast}$ attains its upper limit. Figure 3 exhibits $y_r^{\ast}$ as a function of $E_b$ for various molecular weights $M$. As can be seen, $y_r^{\ast}$ grows more rapidly for larger $M$ and the saturated value for $y_r^{\ast}$ increases with $M$. For example, the saturated limit for $y_r^{\ast}$ grows from $0.023$ to $0.079$ when $M$ is increased from $5$ to $100$. Of course, the saturated value for $y_r^{\ast}$ remains independent of $M$ after $M$ reaches a large value above which most polymer properties become insensitive to changes in molecular weight. Therefore, Figs. 2 and 3 indicate that the influence of chain stiffness becomes more significant for telechelic polymers with weaker sticky interactions and higher molecular weights.

While Figs. 2 and 3 may suggest that the chain stiffness generally promotes self-assembly, Fig. 4 indicates that the model of linear telcehelic polymer melts permits self-assembly also to be suppressed by chain stiffness, a trend that occurs when $\phi$ is sufficiently low at $T=200$ K. Further analysis indicates that temperature also bears on whether chain stiffness promotes or opposes the self-assembly. For example, Fig. 5 displays the contour plot of $y_r^{\ast}$ in the $\phi$-$T$ plane, where the bending energy is fixed to be $E_b=2000$ K. Although chain stiffness promotes self-assembly for systems represented in the $\phi$-$T$ plane where either $\phi$ or $T$ is high (termed region I), self-assembly is indeed suppressed by chain stiffness when both $\phi$ and $T$ are sufficiently low (termed region II). 

The result in Fig. 5 may be rationalized as follows. Linear clusters are expected to form more easily than cyclic clusters (i.e., rings) at high volume fractions, in line with a FH type theory of self-assembly.~\cite{JCP_136_244904} The picture clearly changes when polymer volume fraction is low. In particular, the FH type theory of self-assembly~\cite{JCP_136_244904} predicts that rings predominate over the formed clusters at low temperatures, whereas the opposite situation ensues at high temperatures, a behavior that arises because the extra bond energy gained upon ring closure outweighs the entropy loss upon ring closure as temperature decreases. Therefore, the formation of cyclic clusters is favored only when both $\phi$ and $T$ are low; otherwise, most of the clusters that form in the system are linear. The trend of forming linear clusters is enhanced by chain stiffness because of a diminished probability of ring closure as the chains stiffen. Moreover, the formation of sticky bonds between different chains becomes less sterically hindered as the chains stiffen. Hence, it is observed in Fig. 5 that chain stiffness promotes self-assembly in region I, where linear clusters predominate. Meanwhile, if cyclic clusters predominate, the gain in sticky bonds due to the enhancement of linear clusters induced by chain stiffness cannot compensate for the loss of sticky bonds due to the stiffness generated reduction in ring formation. Consequently, chain stiffness opposes the self-assembly, resulting in the appearance of region II in the $\phi$-$T$ plane in Fig. 5. Clearly, the above explanation is largely based on a FH type theory,~\cite{JCP_136_244904} because the present LCT provides no information concerning the formation of rings. On the other hand, computer simulations may provide a clearer microscopic picture for understanding the trends for the variation of the average degree of self-assembly with chain stiffness.

The presence in Fig. 5 of two regions with opposite behavior implies the existence of a boundary in the $\phi$-$T$ plane that has $y_r^{\ast}=0$ and, therefore, that separates the two regions with opposite dependences of $y^{\ast}$on chain stiffness (see the solid line in Fig. 5). Of course, the concept of a boundary line in the $\phi$-$T$ plane becomes meaningless for linear telechelic polymers with sufficiently large $|\epsilon_s|$, where chain stiffness ceases to affect $y^{\ast}$ (see Fig. 2). An examination of the variation of the boundary with molecular parameters reveals that the boundary line is insensitive to $\epsilon_s$ [Fig. 6(a)]. The area of region II shrinks slightly with increasing $E_b$ and saturates for sufficiently large $E_b$ [Fig. 6(b)], a trend that ensues upon increasing $M$, as shown in Fig. 6(c). Moreover, Figure 6(d) indicates that elevating $\epsilon$ leads to a dramatic increase in the area of region II in the $\phi$-$T$ plane.

\subsection{Influence of chain stiffness on the enthalpy and entropy of self-assembly}

\begin{figure}[tb]
	\centering
	\includegraphics[angle=0,width=0.45\textwidth]{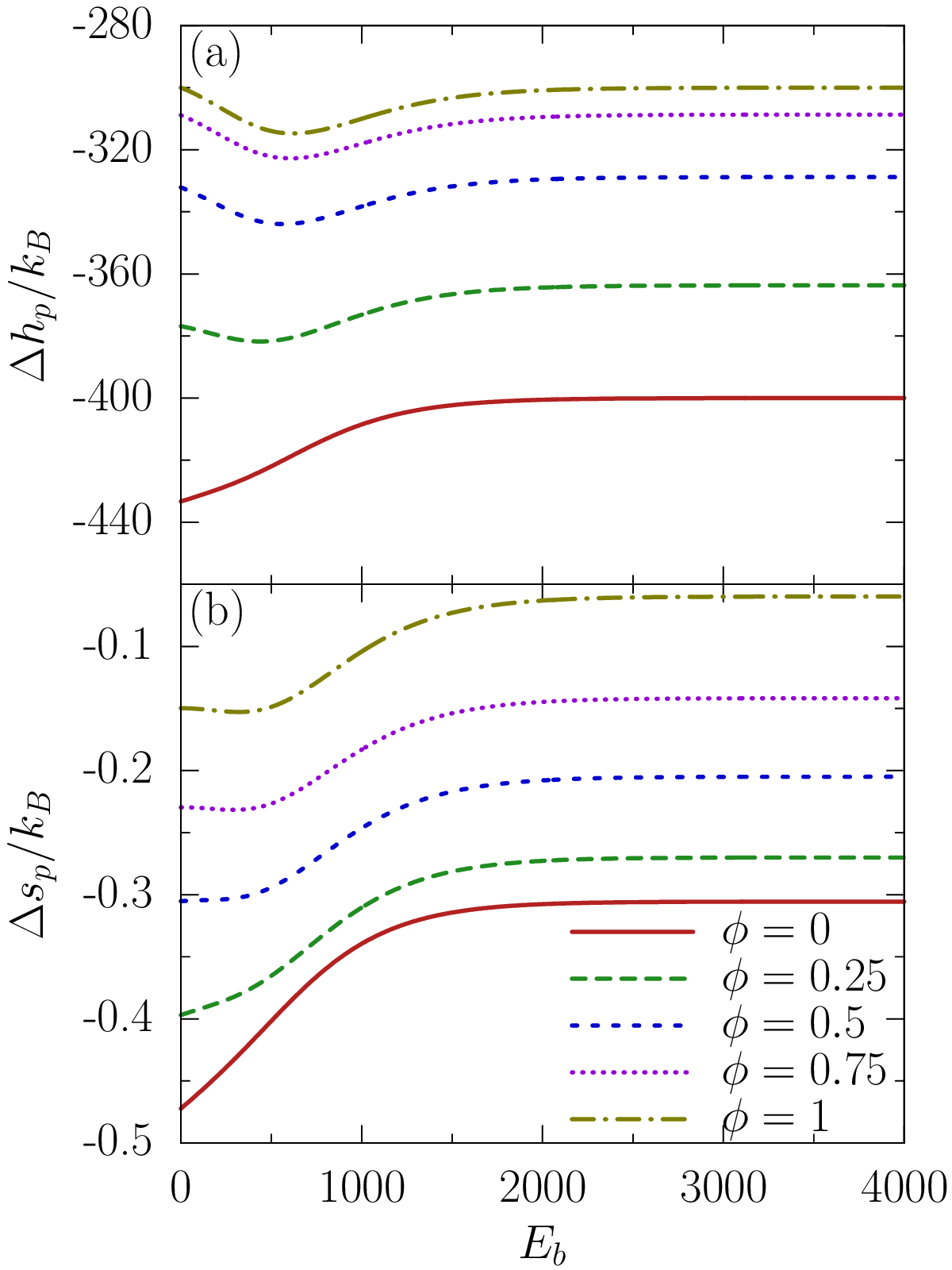}
	\caption{Enthalpy $\Delta h_p$ and entropy $\Delta s_p$ of self-assembly of linear telechelic melts as a function of bending energy $E_b$ for various polymer volume fractions $\phi$. The computations are performed at a constant temperature of $T=300$ K for melts of linear telechelic chains, where the molecular weight is $M=5$, the van der Waals interaction energy is $\epsilon=100$ K, and the sticky interaction energy is $\epsilon_s=-300$ K.}
\end{figure}

The enthalpy $\Delta h_p$ and entropy $\Delta s_p$ of self-assembly characterize the thermodynamics of self-assembly in classical FH type theories for the equilibrium self-assembly of linear polymers,~\cite{JCP_119_12645, JCP_130_164905, JCP_130_224906, JCP_136_244904} but they are commonly treated as phenomenological parameters. On the other hand, the LCT for telechelic polymers allows the determination of $\Delta h_p$ and $\Delta s_p$ as a function of thermodynamic and molecular parameters. The previous analysis~\cite{JCP_136_064903} indicates that $\Delta h_p$ and $\Delta s_p$ strongly depend on various thermodynamic and molecular parameters, and a similar analysis shows that chain stiffness can be added to the list of molecular properties that greatly influence the enthalpy and entropy of self-assembly.

Because the present theory employs a coarse grained model in which each united atom group effectively comprises, say, an actual monomer, changes in enthalpies and entropies on association in real systems may have contributions from the missing internal degrees of freedom that would have to be included as parameters in the present formulation but that would be derived by including these additional degrees of freedom by describing the telechelics with structured monomers. When studying systems with solvent molecules, there may be additional contributions from solvent reorganization, disruption of solvent network, etc. The present theory neglects the contributions arising from the internal degrees of freedom.

The enthalpy $\Delta h_p$ and entropy $\Delta s_p$ of self-assembly are defined as the derivatives of the portion $f_s$ of the free energy $f$ that arises entirely from the sticky interactions,~\cite{JCP_136_064903}
\begin{eqnarray}
\beta \Delta h_p=\beta \frac{1}{y^*}\left.\frac{\partial (\beta f_s)}{\partial \beta}\right|_{N_l,\phi}=\beta \epsilon_s-\sum_{i=1}^{4}(y^*)^{i-1}\beta \frac{\partial Y_i}{\partial \beta}
\end{eqnarray}
and
\begin{eqnarray}
\frac{\Delta s_p}{k_B}&&=\frac{1}{y^*}\left(-\frac{1}{k_B}\left.\frac{\partial f_s}{\partial T}\right|_{N_l,\phi}-\frac{s_{s,comb}}{k_B}\right)\nonumber\\
&&=\sum_{i=1}^{4}(y^*)^{i-1}\left(Y_i-\beta \frac{\partial Y_i}{\partial \beta}\right),
\end{eqnarray}
where $s_{s,comb}$ is the portion of the system's total combinatorial entropy that depends on the density $y^*$ of sticky bonds,
\begin{eqnarray}
\frac{s_{s,comb}}{k_B}=&&\phi x\ln(\phi x)-(\phi x-2y^*)\ln(\phi x-2y^*)\nonumber\\
&&
-y^*\left[1+\ln\left(\frac{2y^*}{z}\right)\right].
\end{eqnarray}

In the limit of infinite dilution (i.e., $\phi=0$ and $\phi=1$), Eqs. (7) and (8) simplify significantly for fully flexible chains,~\cite{JCP_136_064903}
\begin{eqnarray}
\Delta h_p(E_b\rightarrow 0)=\epsilon_s-(1+\frac{2}{z})\epsilon, \text{for $\phi=0$},
\end{eqnarray}

\begin{eqnarray}
\Delta h_p(E_b\rightarrow 0)=\epsilon_s, \text{for $\phi=1$},
\end{eqnarray}

\begin{eqnarray}
\frac{\Delta s_p(E_b\rightarrow 0)}{k_B}=-\frac{2}{z}-\frac{5}{z^2}, \text{for $\phi=0$},
\end{eqnarray}

\begin{eqnarray}
\frac{\Delta s_p(E_b\rightarrow 0)}{k_B}=&&-\frac{1}{z}\frac{2}{M}+\frac{1}{z^2}\left(-3-\frac{4}{M}+\frac{2}{M^2}-\frac{8}{M^3}\right)\nonumber\\
&&
+y^*\left[\frac{1}{z}+\frac{1}{z^2}\left(-\frac{2}{M}+\frac{12}{M^2}\right)\right]\nonumber\\
&&
+(y^*)^2\frac{1}{z^2}\left(\frac{2}{3}-\frac{8}{M}\right)\nonumber\\
&&
+(y^*)^3\frac{2}{z^2}, \text{for $\phi=1$},
\end{eqnarray}

and for fully stiff chains,
\begin{eqnarray}
\Delta h_p(E_b\rightarrow \infty)=\epsilon_s-\epsilon, \text{for $\phi=0$},
\end{eqnarray}

\begin{eqnarray}
\Delta h_p(E_b\rightarrow \infty)=\epsilon_s, \text{for $\phi=1$},
\end{eqnarray}

\begin{eqnarray}
\frac{\Delta s_p(E_b\rightarrow \infty)}{k_B}=-\frac{2}{z}+\frac{1}{z^2}, \text{for $\phi=0$},
\end{eqnarray}

\begin{eqnarray}
\frac{\Delta s_p(E_b\rightarrow \infty)}{k_B}=&&-\frac{1}{z}\frac{2}{M}+\frac{1}{z^2}\left(1-\frac{10}{M}+\frac{18}{M^2}-\frac{8}{M^3}\right)\nonumber\\
&&
+y^*\left[\frac{1}{z}+\frac{1}{z^2}\left(-\frac{10}{M}+\frac{12}{M^2}\right)\right]\nonumber\\
&&
+(y^*)^2\frac{1}{z^2}\left(\frac{2}{3}-\frac{8}{M}\right)\nonumber\\
&&
+(y^*)^3\frac{2}{z^2}, \text{for $\phi=1$}.
\end{eqnarray}

As shown in Eqs. (10-17), the enthalpy $\Delta h_p$ and entropy $\Delta s_p$ of self-assembly display considerable variation even in the limit of infinite dilution when the chains shift from fully flexible to fully stiff. For example, $\Delta h_p$ and $\Delta s_p$ increase by $(2/z)\epsilon$ and $6/z^2$ in the limit of $\phi=0$, respectively, when the bending energy elevates from $E_b\rightarrow0$ to $E_b\rightarrow \infty$. While both limits ($E_b\rightarrow0$ and $E_b\rightarrow\infty$) lead to an identical value of $\Delta h_p=\epsilon_s$ for $\phi=1$, a finite $E_b$ leads to a decrease in $\Delta h_p$ for $\phi=1$ [see Fig. 7(a)]. The entropy $\Delta s_p$ of self-assembly in the limit of $E_b\rightarrow\infty$ is generally found to be larger than that in the limit of $E_b\rightarrow0$. Figure 7 also implies that the dependence of $\Delta h_p$ and $\Delta s_p$ on $E_b$ changes sensitively with other parameters such as $\phi$. 

\subsection{Influence of chain stiffness on the entropy density of linear telechelic melts}

\begin{figure}[tb]
 \centering
 \includegraphics[angle=0,width=0.45\textwidth]{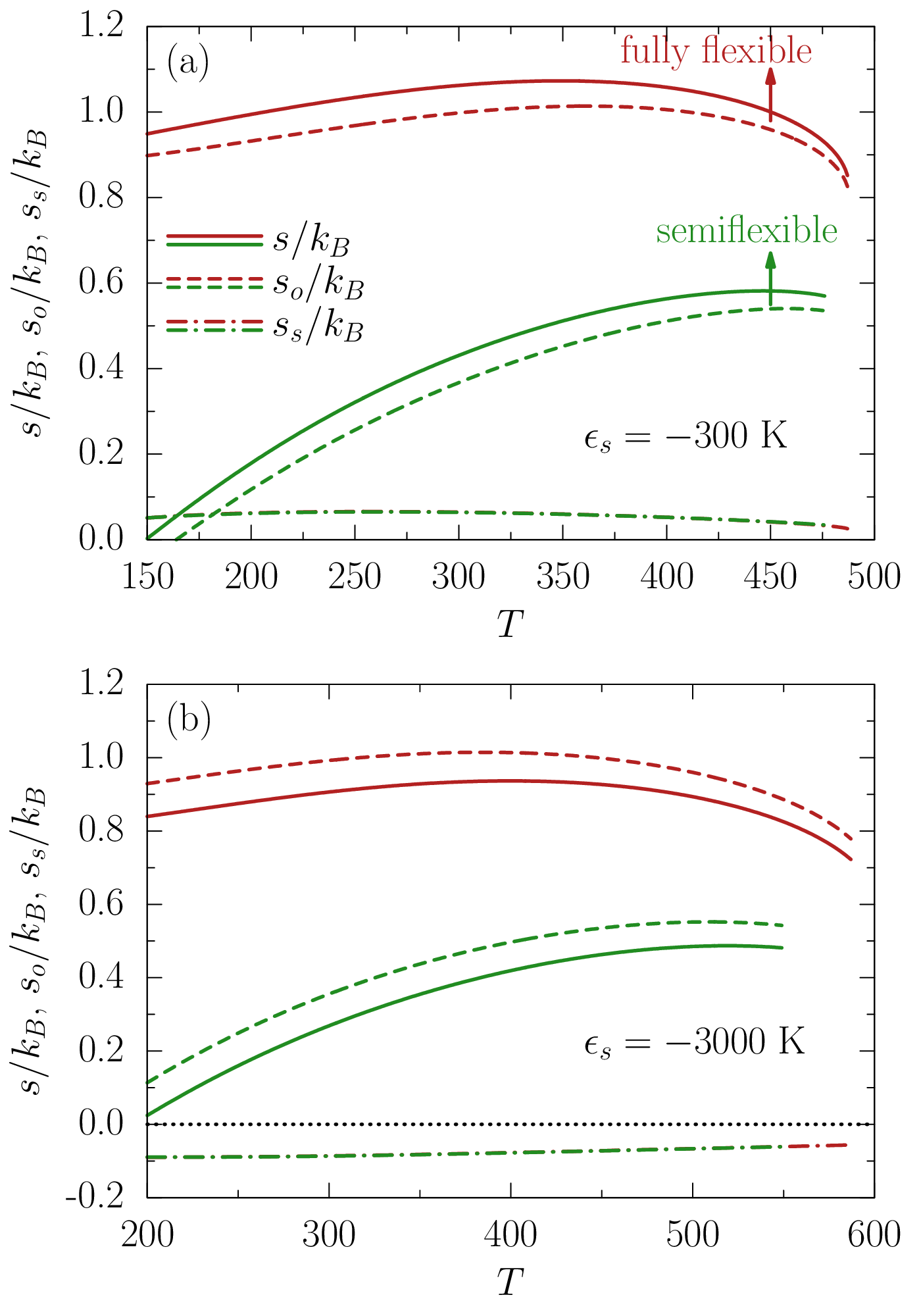}
 \caption{Entropy density $s/k_B$ and its various contributions $s_o/k_B$ and $s_s/k_B$ as a function of temperature $T$ for a melt of fully flexible linear telechelic chains (red lines) and a melt of semiflexible linear telechelic chains (green lines) for (a) a weak sticky interaction strength $\epsilon_s=-300$ K and (b) a strong sticky interaction strength $\epsilon_s=-3000$ K. The computations are performed at a constant pressure of $P=1$ atm. The molecular weight is $M=8$, and the van der Waals interaction energy is $\epsilon=200$ K for both melts, while the bending energy is $E_b=400$ K for the semiflexible melt. The dotted line in (b) highlights the fact that the contribution from the sticky interactions to the entropy density can be negative for very strong sticky interactions.}
\end{figure}

As expected, the chain stiffness of linear telechelic polymer melts exerts a major impact on the entropy density (i.e., the entropy per lattice site), which is a central quantity in the GET of polymer glass formation.~\cite{ACP_137_125} The analytic expression for the entropy density simply follows from the specific Helmholtz free energy $f$ as
\begin{eqnarray}
s=-\left. \frac{\partial f}{\partial T}\right|_{\phi}.
\end{eqnarray}
Likewise, the entropy density is the sum of two separated terms $s_o$ and $s_s$, corresponding to the contributions arising from the reference system and the sticky interactions, respectively. Consequently, $s_o$ and $s_s$ are defined by
\begin{eqnarray}
s_o=-\left. \frac{\partial f_o}{\partial T}\right|_{\phi},
\end{eqnarray}
and
\begin{eqnarray}
s_s=-\left. \frac{\partial f_s}{\partial T}\right|_{\phi}.
\end{eqnarray}
Following previous work,~\cite{ACP_137_125} the computations are performed at constant pressure $P$, which is determined from the Helmholtz free energy $F$,
\begin{eqnarray}
P=-\left.\frac{\partial F}{\partial V}\right|_{m,T}=-\left.\frac{1}{V_{\text{cell}}}\frac{\partial F}{\partial N_l}\right|_{m,T},
\end{eqnarray}
where $V$ is the volume of the system, and $V_{\text{cell}}=a_{\text{cell}}^3$ is the volume associated with a single lattice site. The parameters $z=6$, $a_{\text{cell}}=2.7$\AA{}, and $\epsilon=200$ K are used in our illustrative computations. A low molecular weight of $M=8$ is chosen for the examples because the contribution from the sticky interactions to the entropy density becomes more significant for lower $M$; see discussion below.

Figure 8 displays the temperature dependence of the entropy density $s$ as well as its various contributions $s_o$ and $s_s$ for a melt of semiflexible linear telechelic chains with the bending energy $E_b=400$ K at a constant pressure of $P=1$ atm for weak sticky interactions $\epsilon_s=-300$ K [Fig. 8(a)] and strong sticky interactions $\epsilon_s=-3000$ K [Fig. 8(b)], respectively. The variation for fully flexible linear telechelic polymer melts is also included for comparison, as well as to demonstrate the importance of including chain semiflexibilty within the LCT. (The expression for the specific free energy for fully flexible linear telechelic polymer melts can be found in Ref.~\citenum{JCP_136_064902}.) The term $s_o$ from the reference system dominates the entropy density $s$ of the system for both weak and strong sticky interactions. In fact, the contributions arising from the sticky interactions become even less significant for larger molecular weights, and $s$ is nearly identical to $s_o$ for sufficiently large $M$. This behavior agrees with expectation since the concentration $y^{\ast}$ of sticky bonds becomes extremely small for very large molecular weights.~\cite{JCP_136_064903} Recall that the upper limit for $y^{\ast}$ is inversely proportional to $M$ as $y_{max}^{\ast}=\phi/M$. The above feature applies for both fully flexible and semiflexible linear telechelic polymer melts.

Furthermore, we find that increasing the bending energy has no significant influence on $s_s$ but leads to a dramatic drop in $s_o$ (see either panel of Fig. 8). Consequently, the entropy density $s$ for a semiflexible melt decreases precipitously upon cooling and approaches zero at low temperatures, leading to the appearance of a characteristic feature of glass formation within the CET. (The vanishing of the entropy density is probably an artifact of the high $T$ expansion in the LCT.~\cite{JCP_141_234903}) Therefore, it is possible to explore glass formation in telechelic polymers using the GET only if the polymer chains are modeled as being semiflexible, although the fully flexible telechelic model suffices in establishing the trends obeyed by the thermodynamic properties that are weakly dependent on chain stiffness.~\cite{JCP_136_064903, JCP_136_194902} (The maximum displayed by the entropy density $s$ in Fig. 8 is also a characteristic feature of glass formation within the GET, and Ref.~\citenum{ACP_137_125} discusses why the entropy density depends non-monotonically on temperature.) Figure 8 also reveals that the term $s_s$ arising from the sticky interactions can be either positive or negative, depending on whether the sticky interactions are weak or strong. Thus, it suggests that the sticky interaction energy $\epsilon_s$ may have interesting influences on glass formation in linear telechelic polymer melts. Future work will combine the current extension of the LCT with the AG relation,~\cite{JCP_43_139, JCP_141_141102} to provide a generalization of the GET that enables detailed analysis for the role of self-assembly and the magnitude of the sticky interaction energy $\epsilon_s$ upon glass formation in telechelic polymers.

\section{Summary}

We apply the LCT for semiflexible linear telechelic polymers to assess the influence of chain stiffness on the basic thermodynamic properties of self-assembling telechelic polymer melts and specifically illustrate general trends for the dependence of the average degree of self-assembly upon chain stiffness. The calculations from the LCT imply that chain stiffness promotes self-assembly of linear telechelic polymer melts when either the polymer volume fraction or the temperature is high, but opposes the self-assembly when both polymer volume fraction and temperature are sufficiently low. A FH type theory~\cite{JCP_136_244904} of the competition between the formation of rings versus linear clusters is invoked to provide a possible rationale for the predictions from the LCT concerning the influence of chain stiffness on the average degree of self-assembly. Our results indicate that chain stiffness provides an important molecular variable for tailoring the physical properties of self-assembling telechelic polymers. Meanwhile, we emphasize that simulations clearly offer the possibility of better understanding the influence of chain stiffness on the average degree of self-assembly in telechelic polymers by analyzing the structural and dynamical properties, which are inaccessible by our thermodynamic model. Our theory instead provides a tool for guiding the design of telechelic polymer materials by establishing the relation between the molecular details and the thermodynamic properties.

FH type theories for the equilibrium self-assembly of polymers traditionally employ highly coarse grained models,~\cite{JCP_119_12645, JCP_130_164905, JCP_130_224906, JCP_136_244904} where the interaction parameters (such as the enthalpy and entropy of self-assembly) that characterize the thermodynamics of self-assembly must be adjusted phenomenologically. The LCT for telechelic polymers instead provides a theoretical tool for determining these interaction parameters as a function of molecular and thermodynamic parameters of the self-assembling system. While a previous paper~\cite{JCP_136_064903} illustrates the strong dependence of the enthalpy and entropy of self-assembly on temperature, polymer volume fraction, molecular weight, van der Waals interaction energy, and sticky interaction energy, the present paper continues to demonstrate the significant influence of the bending energy on these interaction variables.

One substantial benefit of the LCT for semiflexible linear telechelic polymers lies in the fact that the characteristic features of glass formation appear in the GET~\cite{ACP_137_125} only if the polymer chains are modeled as being semiflexible. Hence, we provide illustrative calculations for the influence of chain stiffness on the entropy density of linear telechelic polymer melts and demonstrate the importance of including semiflexibility within the LCT for exploring glass formation in linear telechelic polymer melts. Our illustrative calculations also imply that the contribution arising from the sticky interactions to the entropy density can be either positive or negative, depending on whether the sticky interactions are weak or strong, suggesting that the sticky interaction energy may have interesting influences on glass formation in linear telechelic polymer melts. A generalization of the GET for exploring the influence of self-assembly on glass formation in linear telechelic polymer melts can be achieved by combining the current extension of the LCT with the AG relation.~\cite{JCP_43_139, JCP_141_141102}

\begin{acknowledgments}
We thank Jacek Dudowicz for help on the calculations of diagrams with sticky bonds. This work is supported by the National Science Foundation (NSF) Grant No. CHE-1363012.
\end{acknowledgments}

\bibliography{refs}

\end{document}